# Risk of AI in Healthcare: A Comprehensive Literature Review and Study Framework

## Apoorva Muley [a], Prathamesh Muzumdar [b*], George Kurian [c] and Ganga Prasad Basyal [d]

[a] *People's University, Bhopal, India.*
[b] *The University of Texas at Arlington, USA.*
[c] *Eastern New Mexico University, USA.*
[d] *Black Hills State University, USA.*

*Review Article*

## ABSTRACT

This study conducts a thorough examination of the research stream focusing on AI risks in healthcare, aiming to explore the distinct genres within this domain. A selection criterion was employed to carefully analyze 39 articles to identify three primary genres of AI risks prevalent in healthcare: clinical data risks, technical risks, and socio-ethical risks. Selection criteria was based on journal ranking and impact factor. The research seeks to provide a valuable resource for future healthcare researchers, furnishing them with a comprehensive understanding of the complex challenges posed by AI implementation in healthcare settings. By categorizing and elucidating these genres, the study aims to facilitate the development of empirical qualitative and quantitative research, fostering evidence-based approaches to address AI-related risks in healthcare effectively. This endeavor contributes to building a robust knowledge base that can inform the

_________________________________________________________________________________

*Corresponding author: E-mail: prathameshmuzumdar85@gmail.com;*

formulation of risk mitigation strategies, ensuring safe and efficient integration of AI technologies in healthcare practices. Thus, it is important to study AI risks in healthcare to build better and efficient AI systems and mitigate risks.

## 1. INTRODUCTION

Some individuals and organizations argue that the overall potential of AI in medicine has been largely overestimated, as there is a lack of concrete data demonstrating significant improvements in patient outcomes. This viewpoint raises skepticism about the widespread adoption and transformative power of medical AI. Additionally, experts have voiced concerns in recent years regarding potential adverse consequences of medical AI [1]. These concerns encompass various aspects, including clinical, technical, and socio-ethical risks. Such issues highlight the need for careful evaluation and regulation of AI technologies in the healthcare domain to ensure patient safety and to address any unintended negative impacts.

While AI in healthcare holds promise and continues to advance, it is essential to critically assess its performance, potential benefits, and risks to make informed decisions about its integration into medical practice. As with any emerging technology, cautious and responsible implementation is key to unlocking its true potential while minimizing potential downsides [2]. In the literature, several main risks and challenges have been identified as likely to arise from the introduction of AI in future healthcare. These risks and challenges can be categorized into seven major categories:

1. Patient harm due to AI errors, 2. Misuse of medical AI tools, 3. Risk of bias in medical AI and perpetuation of inequities, 4. Lack of transparency, 5. Privacy and security issues, 6. Gaps in AI accountability, 7. Obstacles to implementation in real-world healthcare

Addressing these risks requires close collaboration between various stakeholders, including healthcare professionals, AI developers, policymakers, and ethicists. Ensuring robust evaluation, regulation, and continuous monitoring of AI systems are crucial to maximize the benefits of AI while minimizing potential negative impacts on patient care and healthcare delivery [3]. The study has reviewed 39 articles from last 5 years from 2018 to 2023 and come up with a risk of AI in healthcare study framework incorporating the 7 major categories of AI risk. This study deep dives in each category to help readers understand the findings from the reviewed literature [4]. Overall, the study helps to summarize the recent finding and develop a review study framework. The aim of this study is to provide a clear and coherent framework for understanding and mitigating the risks of AI in healthcare.

## 2. METHODS FOR COMPREHENSIVE REVIEW

This study represents a pioneering effort, employing a 5-year retrospective literature analysis of research papers addressing the topic of AI risks in healthcare. To select the relevant articles, the criteria focused on the presence of the term "Risk of AI in healthcare/medical science/medicine/biomedical science" in the title, abstract, or keywords. Through this meticulous search process, a total of 39 journal articles were identified, subsequently read, coded, and categorized according to specific themes and classifications detailed in Table 1 from the Appendix.

While a majority of the research on AI risks in healthcare explicitly adopts a theoretical framework, these frameworks primarily rely on real-world medical cases derived from medical data. To better comprehend the diverse range of AI risks, the study classified them into three main genres: clinical data risks, technical risks, and socio-ethical risks, as illustrated in Fig. 1. Moreover, each genre was further sub-categorized to facilitate a comprehensive understanding and examination of AI risks, given their variability across different scenarios. The study delves deeply into each genre of AI risks, providing a detailed and insightful explanation of their implications and significance in the healthcare context.

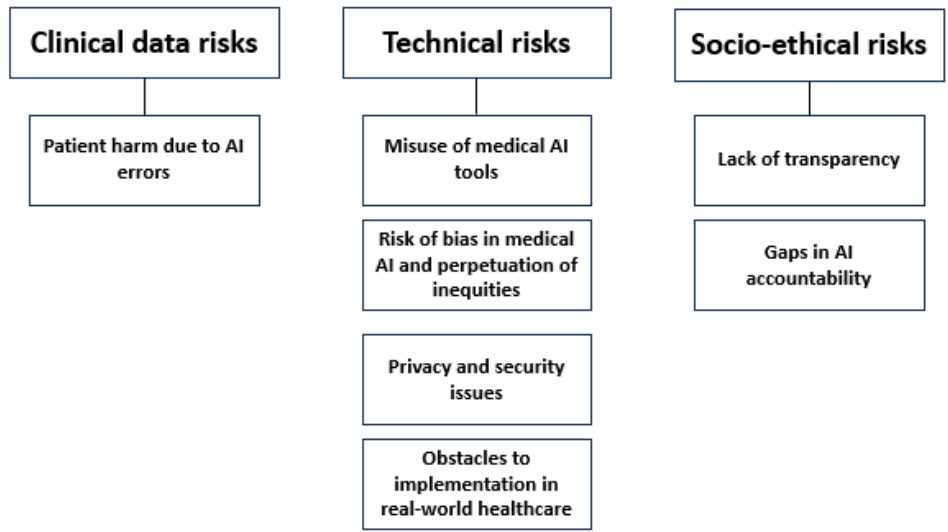

**Fig. 1. Risk of AI in healthcare literature review study framework**

## 3. CLINICAL RISKS

### 3.1 Patient Harm Due to AI Errors

In spite of continuous advancements in data availability and machine learning, AI-driven clinical solutions in healthcare may lead to failures, posing potential safety concerns for healthcare service users. Even with AI developers having access to extensive, high-quality datasets for training their AI technologies, there remain at least three major sources of error in AI implementation within clinical practice [22]. Firstly, the accuracy of AI predictions can be significantly affected by noise in the input data while using the AI tool [6]. Secondly, AI misclassifications may occur due to dataset shift, a common machine learning issue where the statistical distribution of data used in clinical practice deviates, even slightly, from the original dataset used for AI training [31][1]. This shift could arise from differences in population groups, acquisition protocols among hospitals, or the use of machines from various manufacturers [39]. For instance, a multi-center study in the United States developed a highly accurate AI system for diagnosing pneumonia based on data from two hospitals [3]. However, when tested with data from a third hospital, a considerable decrease in accuracy was observed, indicating potential hospital-specific biases.

Lastly, predictions can be prone to errors because AI algorithms struggle to adapt to unexpected changes in the environment and context of their application. For example, in medical imaging, the AI model might mistake regular artifacts as observational errors, leading to false positives.

### 3.2 Remedies to Prevent Patient Harm due to AI Errors

To ensure the safe and effective use of AI solutions in healthcare, several key steps must be taken. First and foremost, standardized methods and procedures should be established for extensive evaluation and regulatory approval of AI technologies [40]. This evaluation process should specifically assess the ability of AI solutions to generalize to new populations and their sensitivity to noise. Secondly, AI algorithms should be designed and implemented as assistive tools rather than fully autonomous systems [40]. This means that clinicians should remain an integral part of the data processing workflow, enabling them to detect and report potential errors and contextual changes. By doing so, the aim is to minimize any potential harm to patients.

In addition to being designed as assistive tools, future AI solutions in healthcare must be dynamic. They should be equipped with mechanisms that enable them to continuously learn from new scenarios and mistakes detected in real-world practice [40]. However, it's important to strike a balance and maintain a certain degree of human control and vigilance to promptly identify and address any emerging problems.

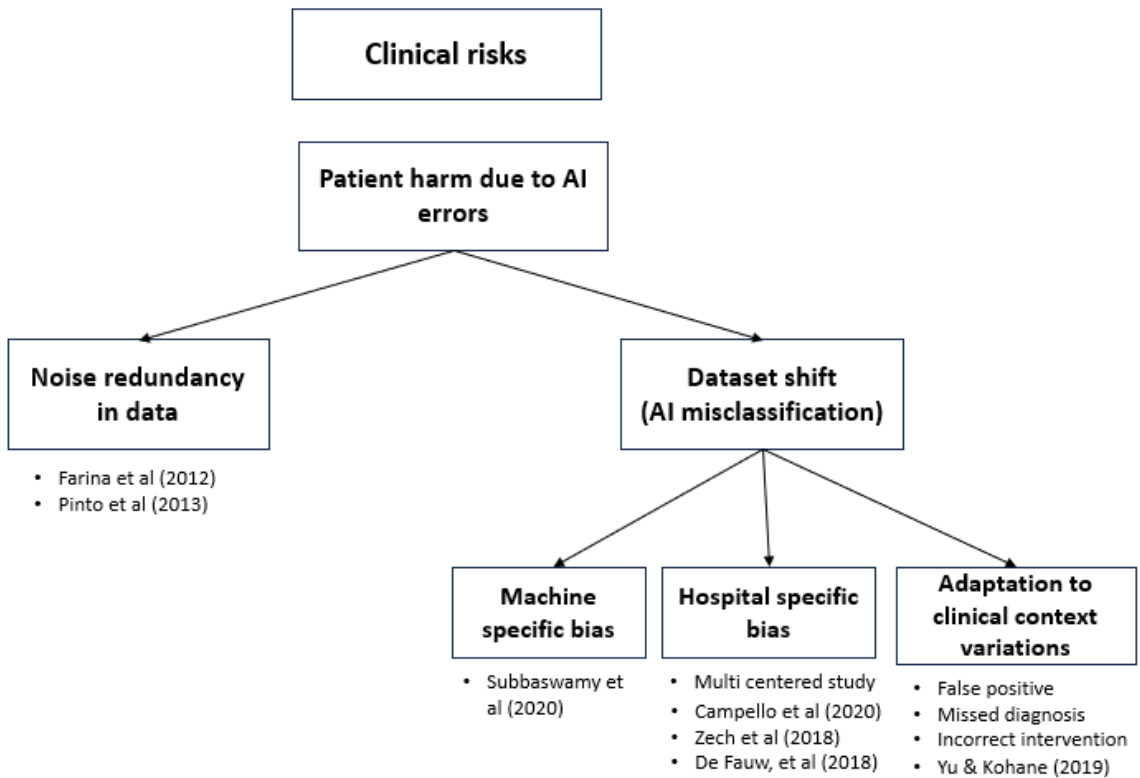

**Fig. 2. Patient harm due to AI errors**

This may lead to increased costs and initially reduce some of the immediate benefits of AI implementation. To support these dynamic AI systems, infrastructural and technical developments will be necessary to facilitate regular AI updates based on past and new training data [40]. Moreover, policies should be implemented to ensure seamless integration of these mechanisms into healthcare settings, thereby fostering a safer and more efficient use of AI in the medical field.

## 4. TECHNICAL RISKS

### 4.1 Misuse of Medical AI Tools

Like any health technology, medical AI carries the risk of human error and misuse. Even when AI algorithms are accurate and robust, their proper use in practice by end-users, including clinicians, healthcare professionals, and patients, is crucial [28]. Incorrect usage can lead to erroneous medical assessments and decisions, potentially harming the patient. Therefore, it is not sufficient for clinicians and the public to have access to medical AI tools; they must also understand how and when to use these technologies. Several factors make existing medical AI technologies susceptible to human error or incorrect use [32]. Often, these technologies have been designed by computer/data scientists with limited involvement from end-users and clinical experts. Consequently, users, such as clinicians, nurses, data managers, or patients, are required to learn and adapt to the new AI technology, leading to complex interactions and experiences. This complexity can hinder the effective application of AI algorithms in day-to-day clinical practice, reducing the potential for informed decision-making and increasing the likelihood of human error.

These concerns about AI education and literacy also extend to citizens and patients who will use future medical AI solutions. Another potential cause for misuse of medical AI, resulting in harm to citizens and patients, is the proliferation of easily accessible AI applications [24]. While such tools offer convenient options for remote diagnosis and disease monitoring, there is often limited information about how the AI algorithms have been developed and validated, and their reliability and clinical efficacy may not be adequately demonstrated. This situation is reminiscent of easily accessible online pharmacies contributing to medication abuse by citizens, raising public health concerns. Since the development and commercialization of AI-powered web/mobile health applications offer significant financial opportunities, this sector attracts many players and companies with varying standards of ethics, excellence, and quality. While companies offering these AI medical tools may include disclaimers on their websites, users may not always come across, read, or fully understand these disclaimers. As a result, they might rely on potentially incorrect information and diagnoses provided by the AI tools, adversely impacting their health-related decision-making process.

#### 4.1.1 Remedies to prevent misuse of AI medical tools

To ensure the successful and responsible integration of AI in healthcare, it is essential to involve end-users such as healthcare professionals, specialists, technicians, and patients closely in the design and development of AI solutions. This collaborative approach ensures that their perspectives, preferences, and real-world contexts are integrated into the final tools that will be deployed and used. Moreover, it is crucial to develop and generalize education and literacy programs on AI and medical AI across education circles and society. These programs will enhance the knowledge and skills of future AI end-users, reducing the likelihood of human error and promoting the effective use of AI technologies in healthcare. Public agencies should also play a vital role in regulating the sector of web/mobile medical AI [40]. Through proper regulation and oversight, citizens can be well-informed and protected against the misuse and abuse of easily accessible AI technologies. Transparent guidelines and standards are necessary to maintain public trust and safeguard patient interests in the rapidly evolving field of AI in healthcare.

### 4.2 Risk of Bias in Medical AI and Perpetuation of Inequities

Despite significant advancements in medical research and healthcare delivery, inequalities and inequities persist within most countries worldwide. These disparities are influenced by various factors, including sex/gender, age, ethnicity, income, education, and geography [9]. Some of these inequities stem from systemic issues, such as socioeconomic differences and discrimination, while human biases also contribute significantly [27]. An illustrative example of bias is evident in research conducted in the United States, which revealed that doctors may not take complaints of pain from Black patients as seriously or respond to them as quickly as they do for their White counterparts. This disparity persists in varying degrees in many countries worldwide. Another example of common bias embedded in healthcare systems is gender-based discrimination [20]. Studies in pain management have highlighted how female patients' reports of pain may be psychologized or disregarded, indicating a concerning gender bias.

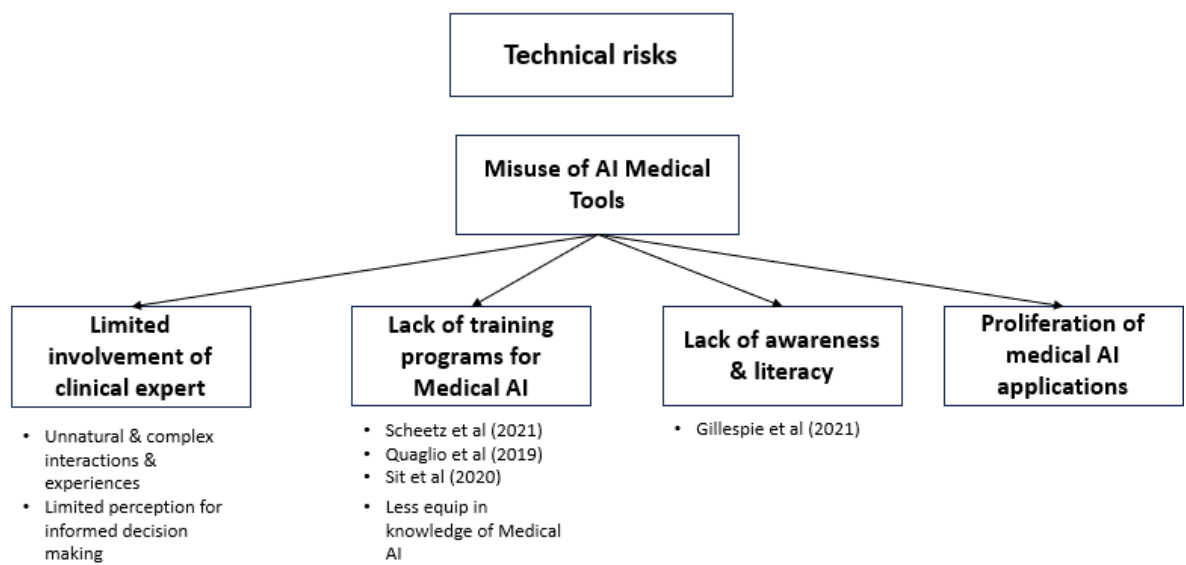

**Fig. 3. Misuse of AI medical tools**

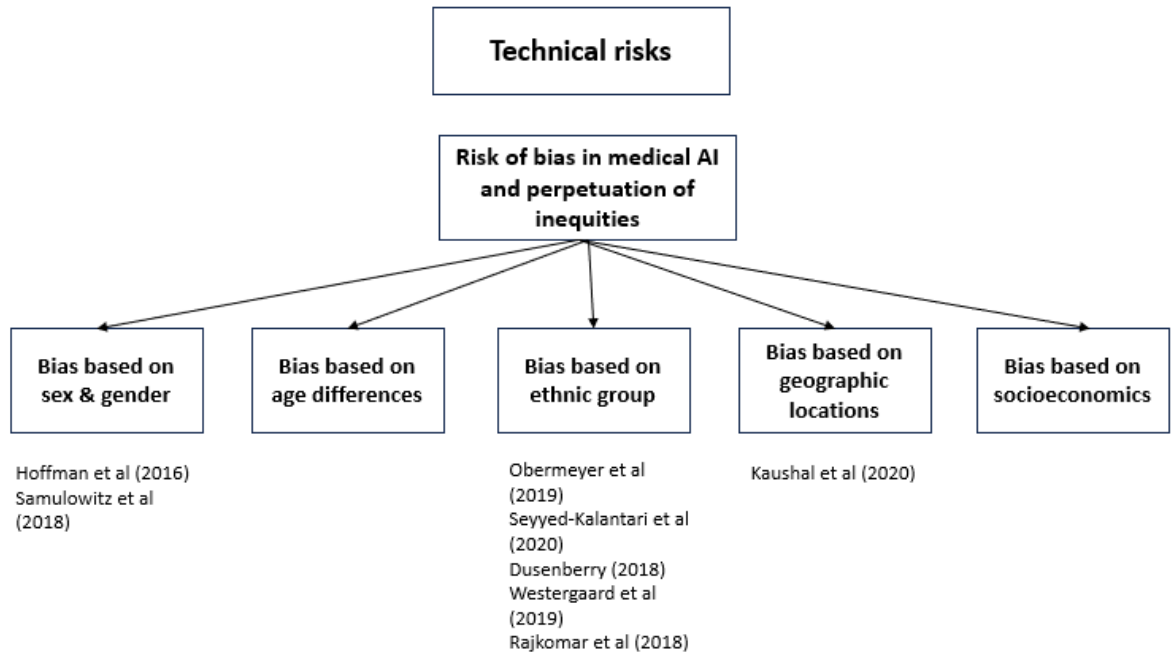

**Fig. 4. Risk of Bias**

These biases and inequalities in medical care underscore the importance of addressing systemic issues and human biases in healthcare systems globally [29]. By recognizing and actively working to eliminate these disparities, we can strive towards a more equitable and just healthcare environment for all individuals, regardless of their background or characteristics. There are growing concerns that if not implemented, evaluated, and regulated properly, future AI solutions could embed and possibly amplify the systemic disparities and human biases contributing to healthcare inequities. Several examples of algorithmic biases have already received significant attention in recent years, some of which are outlined below.

One prominent form of bias is algorithmic bias based on demographic factors such as race, gender, or age [35]. For instance, certain AI algorithms might inadvertently discriminate against specific population groups due to biased data used for training. Geographic bias is another type of bias that appears in datasets [26]. It occurs when AI models are trained on data that primarily represent specific regions or healthcare environments, leading to disparities in care for individuals from other geographic areas. Moreover, bias in data labeling during clinical assessment can impact AI model training and predictions. If certain groups have historically been misdiagnosed or stigmatized, this bias can be reflected in the data used to train AI models, potentially perpetuating disparities in healthcare.

Interestingly, some healthcare conditions, such as injuries, poisonings, congenital malformations, and infectious diseases, exhibit discrepancies that cannot be explained by anatomical or genetic differences. If health registries' data labels are affected by these disparities, AI models might inadvertently learn and perpetuate the biases present in the data [11]. To address these concerns, it is crucial to take steps towards ensuring fairness, transparency, and equity in AI development and deployment in healthcare. Robust evaluation of AI algorithms for bias and the establishment of ethical guidelines and regulations can help mitigate these potential biases and promote more equitable healthcare outcomes for all individuals.

**4.2.1 Remedies to prevent risk of bias in medical AI and perpetuation of inequities**

To ensure fairness and reduce biases in AI development for healthcare, collaboration between AI developers, clinical experts, healthcare professionals, and various stakeholders is essential. Data selection and labelling should be thoughtfully done, with a focus on representing diverse attributes like sex/gender, age, socioeconomics, ethnicity, and geographic location. Building interdisciplinary development teams that include data scientists, biomedical researchers, social scientists, biomedical ethicists, public health experts, patients, and citizens can result in more inclusive and equitable AI tools [40]. Adequate representation of diverse backgrounds and needs through community engagement fosters the creation of AI solutions that genuinely serve the population they are designed for.

Transparency and explainability in AI models enable developers to understand the reasoning behind the model's decisions, making it easier to identify and address potential biases [40]. Continuous monitoring and evaluation of AI models in real-world healthcare settings are necessary to identify and rectify any biases that may arise over time, ensuring the AI tools remain fair and unbiased. By adhering to these principles, AI developers can work towards creating healthcare AI solutions that are more representative, fair, and equitable, contributing to the advancement of inclusive healthcare practices.

### 4.3 Privacy and Security Issues

The increasing development of AI solutions and technology in healthcare, particularly evident during the COVID-19 pandemic, brings potential risks to data privacy, confidentiality, and patient and citizen protection [21]. These risks include the exposure and misuse of sensitive data, which may violate individual rights and lead to non-medical use of patient data. A critical aspect of these issues is informed consent, which involves providing patients with sufficient information to make informed decisions about sharing their personal health data. Informed consent is integral to the patient's healthcare experience and is rooted in ethical principles like protection from harm, respect for autonomy, and privacy rights [23].

However, the introduction of opaque AI algorithms and complex informed consent forms can limit patient autonomy and shared decision-making with physicians [34]. Patients may struggle to understand the data-sharing process and their options for opting out. Big data research and digital platform-based health data research pose additional challenges as patients may not fully comprehend the extent of data sharing and reuse [16][13]. Moreover, the use of AI in healthcare introduces data security risks, with potential breaches leading to privacy violations and identity theft. Cyberattacks on AI systems and personal medical devices controlled by AI also pose serious concerns, highlighting vulnerabilities in the technology [10][12].

Addressing these issues requires comprehensive efforts to enhance transparency, provide clear and accessible information to patients, and strengthen data security measures [19]. Ethical considerations and robust safeguards are essential to ensure the responsible and secure integration of AI in healthcare while safeguarding patient rights and privacy.

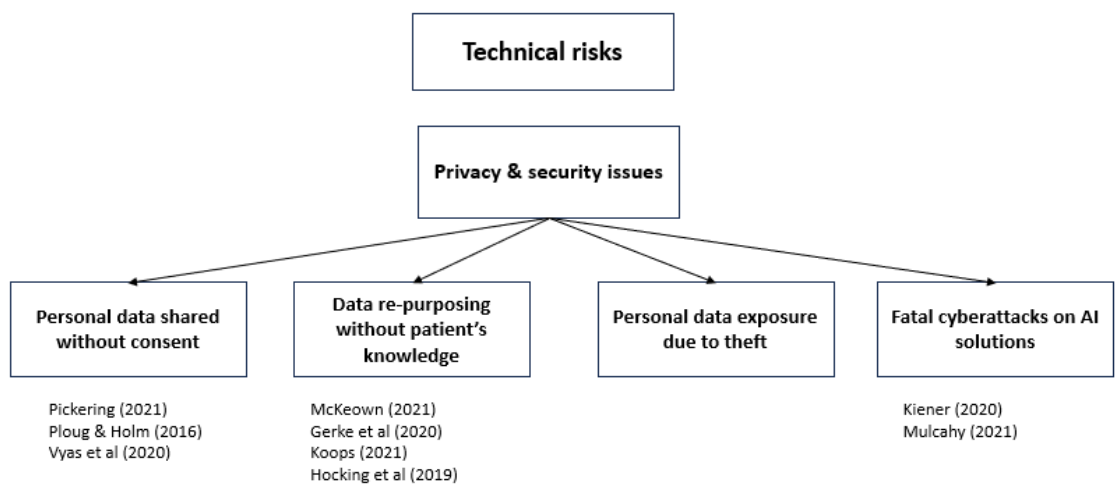

**Fig. 5. Privacy & security issues**

#### 4.3.1 Remedies to privacy and security issues

To tackle these critical issues, it is crucial to increase awareness and knowledge about privacy and security risks, informed consent, and cybersecurity. Additionally, regulatory frameworks should be expanded to address accountability and protect individuals from data breaches and data repurposing. Encouraging decentralized and federated approaches to AI can leverage clinical data without compromising its security [40]. Research should be prioritized to enhance security in cloud-based systems and protect AI algorithms from cyberattacks, ensuring the safe and responsible use of AI in healthcare. Collaboration among stakeholders, including researchers, policymakers, and healthcare professionals, is essential to address these challenges and create a trustworthy AI-driven healthcare environment.

### 4.4 Obstacles to Implementation in Real-World Healthcare

Over the past five years, numerous medical AI algorithms have been developed and proposed, covering a wide range of medical applications. Despite being well-validated, clinically robust, safe, and ethically compliant, the implementation and integration of medical AI technologies in healthcare face various challenges in the real world. Healthcare professionals have traditionally been slower to adopt new technologies compared to other fields. The implementation phase is a crucial stage in the innovation process, and it requires more than just inventing and testing the AI technology.

One significant obstacle to medical AI implementation is the quality of electronic health data in real-world practice. Medical data is often unstructured and noisy, and existing datasets may not be suitable for AI algorithms. Moreover, different clinical centers and EU member states may have varying formats and data quality, necessitating costly human revision, quality control, cleaning, and relabeling before AI tools can be effectively used on a large scale. To address this, efforts should be made to promote better re-use of diverse health data types, such as electronic health records, genomics data, and patient registries across EU countries, thereby benefiting emerging AI algorithms.

AI technologies also have the potential to alter the dynamics of the clinician-patient relationship in ways that are not yet fully understood. Communicating information about AI-derived risks of developing illnesses, like cancer or dementia predisposition, raises personal and ethical implications that need careful consideration. Updating clinical guidelines and care models will be necessary to account for the AI-mediated interactions between healthcare workers and patients, as the introduction of AI technology into everyday practice will have practical, technical, and clinical implications for both clinicians and patients.

### 4.5 Lack of Transparency

Despite significant advancements in medical AI, many individuals and experts still perceive existing algorithms as complex and difficult to comprehend, leading to challenges in fully trusting and adopting these technologies [37]. The lack of transparency is a prominent issue in the development and use of current AI tools in healthcare. This lack of transparency is particularly concerning in sensitive areas like medicine and healthcare, where the well-being and health of citizens are at stake [17]. As a consequence, there is a considerable lack of trustworthiness associated with AI, especially in the medical domain.

The limited trustworthiness is expected to have a significant impact on the adoption of emerging AI algorithms by patients, clinicians, and healthcare systems [30]. AI transparency is closely associated with traceability and explainability. These concepts represent two distinct levels of transparency required in AI applications. First, there is the transparency of the AI development and usage processes, which refers to traceability [38]. Second, there is the transparency of AI decisions, known as explainability. Both traceability and explainability are crucial in instilling trust and confidence in AI systems used in healthcare.

Traceability is a crucial aspect of ensuring trustworthy AI systems [14]. It involves transparently documenting the entire AI development process, from its creation to its real-world performance after deployment. This includes tracking various aspects:

1. Model Details: Documenting specific information about the AI model, such as its intended use, type of algorithm or neural network, hyper-parameters, and any pre- and post-processing steps applied [25].

2. Training and Validation Data: Maintaining a complete account of the data used to train and validate the AI model, including the data gathering process, data composition, acquisition protocols, and data labelling methods [33].
3. AI Tool Monitoring: Keeping track of AI tool performance metrics, instances of failures, and periodic evaluations to assess its effectiveness and potential limitations [36].

Unfortunately, in the practical implementation of existing AI tools in healthcare, full traceability is often lacking. Companies may choose not to disclose comprehensive information about their algorithms, leading to the delivery of opaque tools that are difficult for independent parties to understand and examine. This lack of transparency reduces the level of trust in these AI tools and hinders their adoption into real-world practice. Enhanced traceability is necessary to build confidence and foster broader acceptance of AI in healthcare by ensuring transparency, accountability, and reliability.

AI explainability plays a critical role in providing transparency for each AI prediction and decision. However, some AI solutions, particularly deep neural networks, lack transparency and are often referred to as 'black box AI.' This term reflects the complexity of these models, which learn intricate functions that are challenging for humans to comprehend, making their decision-making processes invisible and difficult to understand [18][8]. The absence of transparency poses significant challenges for clinicians and other stakeholders when incorporating AI solutions into real-world practice. Clinicians need to understand the underlying principles behind each AI decision or prediction to confidently work with specific AI tools, even if these algorithms have the potential to enhance their productivity.

Moreover, the lack of explainability hampers the ability to pinpoint the source of AI errors and assign responsibilities when issues arise. Identifying the root cause of errors becomes challenging due to the opacity of the AI models. To address these limitations, efforts are being made to develop AI explainability techniques that shed light on the decision-making processes of complex AI models. Explainable AI is critical to build trust, facilitate integration into real-world practice, and enhance accountability in the deployment of AI solutions in various domains, including healthcare.

### 4.5.1 Remedies to prevent lack of transparency

Several avenues are available to enhance the transparency of AI technologies in healthcare. Firstly, implementing an 'AI passport' for each AI algorithm can document essential information about the model, promoting understanding and transparency [40]. Secondly, developing traceability tools would enable monitoring the AI algorithms' usage after deployment, recording errors, performance degradation, and facilitating periodic audits. Thirdly, involving clinical end-users from the beginning of the development process would ensure the selection of appropriate explainability approaches and acceptance in clinical practice. Lastly, regulatory entities can encourage transparency by considering traceability and explainability as prerequisites for certification, fostering accountability and informed decision-making. These efforts aim to create more transparent and trustworthy AI solutions in healthcare.

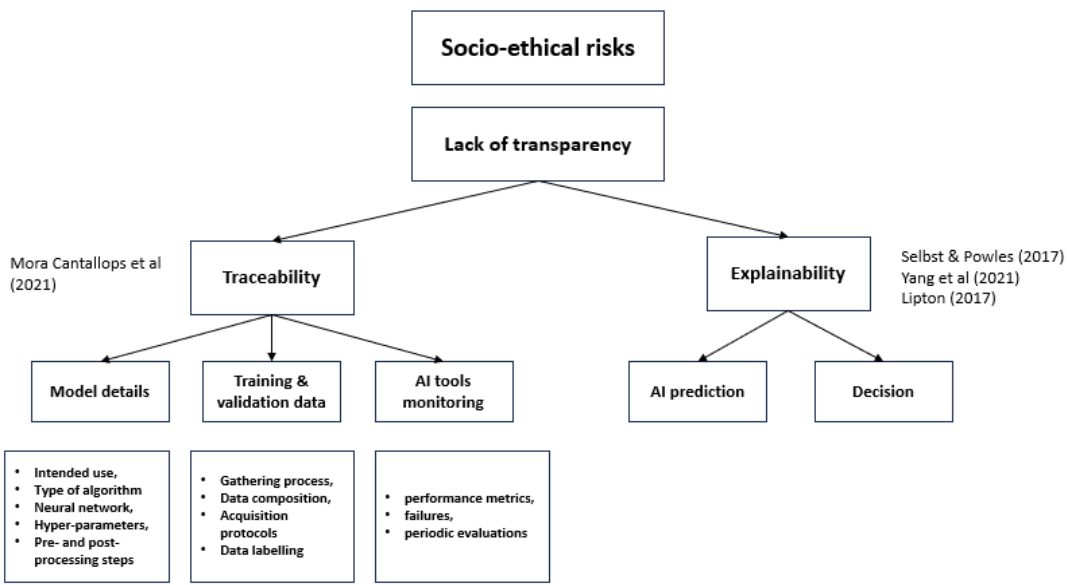

**Fig. 6. Lack of transparency**

### 4.6 Gaps in AI Accountability

The term 'algorithmic accountability' has gained significance in addressing the legal implications of AI algorithms' introduction and use in various aspects of human life [25]. Contrary to what the term might imply, it emphasizes that algorithms are a combination of machine learning and human design, and any errors or wrongdoings originate from the humans involved in their development, introduction, or use. AI systems themselves cannot be held morally or legally responsible. In the context of medical AI, accountability is crucial for its acceptance, trustworthiness, and future adoption in healthcare. If clinicians are held responsible for AI-related medical errors despite not designing the algorithms, they may be hesitant to adopt these AI solutions [33]. Similarly, patients and citizens may lose trust if they believe no one can be held accountable for the potential harm caused by AI tools.

The novel nature of medical AI and the lack of legal precedence led to ambiguity regarding responsibilities for AI-related medical errors that could harm patients. The rapidly evolving field of medical AI poses challenges for regulators, policymakers, and legislators, requiring adaptations to address accountability and liability in AI-assisted healthcare. Challenges in applying current law and liability principles to emerging AI applications in medicine include the involvement of multiple actors, making it difficult to identify responsibilities among AI developers, data managers, clinicians, patients, healthcare organizers, etc. Additionally, determining the precise cause of AI-related medical errors is challenging, as it can stem from the algorithm, training data, or incorrect use in clinical practice [36]. Moreover, the existence of multiple governance frameworks and the absence of unified ethical and legal standards in AI industries add complexity to the situation.

The introduction of AI tools in healthcare expands the patient-clinician dynamic, involving various actors like AI developers, researchers, and manufacturers in medical decision-making [18]. This complexity further contributes to the challenge of assigning accountability. While medical professionals are under regulatory responsibility and may lose their license for not being able to account for their actions, AI developers and technologists typically adhere to ethical codes [8]. The vagueness and enforceability of these codes have raised criticisms. In conclusion, addressing algorithmic accountability in medical AI requires clear regulations, unified standards, and transparent accountability frameworks involving all stakeholders to ensure responsible and trustworthy deployment of AI in healthcare.

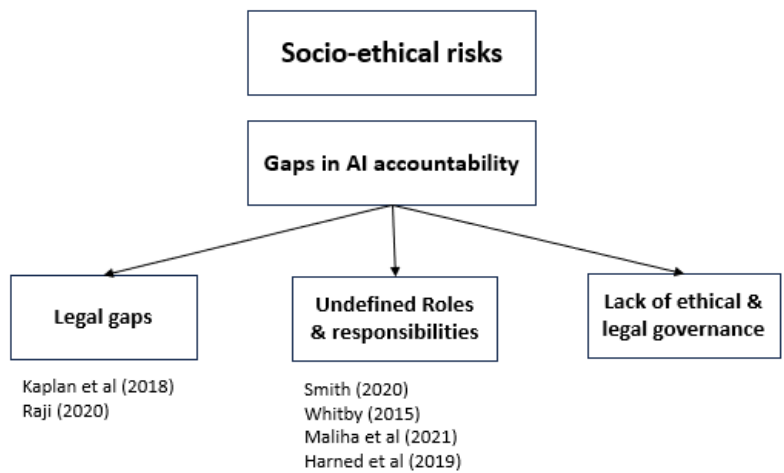

**Fig. 7. Gaps in AI accountability**

#### 4.6.1 Remedies to prevent gaps in AI accountability

To address the current lack of accountability in medical AI, it is important to establish processes for identifying the roles of AI developers and clinical users when AI-assisted medical decisions result in harm to individuals. Additionally, creating dedicated regulatory agencies for medical AI can help develop and enforce frameworks that hold specific actors, including AI manufacturers, accountable for their actions. These measures will promote transparency, trust, and responsible practices in the use of AI in healthcare [40].

## 5. CONCLUSION

In conclusion, this comprehensive literature review has meticulously examined 39 articles focusing on the risks of AI in healthcare. Through a systematic analysis of the literature, a robust framework has been developed, shedding light on three primary genres of AI risks: clinical data risks, technical risks, and socio-ethical risks. Delving deeper into each genre, the study explores various sub-genres, offering a nuanced understanding of the multifaceted challenges surrounding the implementation of AI in healthcare. By providing this detailed reference base, the article equips researchers, policymakers, and healthcare practitioners with valuable insights to foster empirical qualitative and quantitative research in the domain of AI risks in healthcare. This framework not only enhances our comprehension of the potential pitfalls associated with AI adoption but also serves as a crucial guide in designing effective risk mitigation strategies.

As AI continues to evolve and permeate healthcare settings, it is imperative to acknowledge and address the inherent risks involved. This literature review study significantly contributes to the ongoing discourse on AI's role in healthcare by facilitating evidence-based decision-making, ensuring the responsible and safe integration of AI technologies in the healthcare landscape. Moving forward, the findings of this review can serve as a stepping stone for future investigations, urging scholars to delve deeper into specific sub-genres and explore emerging risks that may arise as AI applications advance. As technology progresses, so too should our understanding of the potential hazards and opportunities that AI presents in healthcare. By building upon this comprehensive study, the researchers in AI healthcare can foster a more sustainable and patient-centric AI-driven healthcare system.

## CONSENT

It is not applicable.

## ETHICAL APPROVAL

It is not applicable.

## COMPETING INTERESTS

Authors have declared that no competing interests exist.

# APPENDIX

## Table 1. List of review articles

| No | Authors | Risk Theme | Sub-Theme | Focus | Journal |
|---|---|---|---|---|---|
| 1 | Gerke et. al (2020) | Socio-ethical risk | Lack of transparency | | Book chapter: Artificial intelligence in healthcare |
| 2 | Challen et. al (2019) | Clinical data risk | Dataset shift | | BMJ Quality & Safety |
| 3 | Ellahham et. al (2020) | General risk | General theory | | American Journal of Medical Quality |
| 4 | Manne & Kantheti (2021) | General risk | General theory | | Current Journal of Applied Science and Technology |
| 5 | Pinto et. al (2013) | Clinical data risk | Noise | | Critical Ultrasound Journal |
| 6 | Farina et. al, (2012) | Clinical data risk | Noise | | Errors in radiology |
| 7 | Subbaswamy et. al (2020) | Clinical data risk | Dataset shift | General misclassification | Biostatistics |
| 8 | Campello et. al 2020 | Clinical data risk | Dataset shift | Machine specific bias | Medical Image Computing and Computer Assisted Intervention |
| 9 | Zech et al (2018) | Clinical data risk | Dataset shift | Hospital-specific bias | PLoS Medicine |
| 10 | De Fauw et al (2018) | Clinical data risk | Dataset shift | Hospital-specific bias | Nature medicine |
| 11 | Scheetz et al (2021) | Technical risk | Training programs | Professional training programs | Scientific Reports |
| 12 | Sit et al (2020) | Technical risk | Training programs | Curriculum based formal education | Insights into Imaging |
| 13 | Quaglio et al (2019) | Technical risk | Training programs | General | EPRS, European Parliament |
| 14 | Hoffman et al (2016) | Technical risk | Bias in AI | Gender bias | Proceedings of the National Academy of Sciences |
| 15 | Samulowitz et al (2018) | Technical risk | Bias in AI | Gender bias | Pain Research and Management |
| 16 | Obermeyer et al (2019) | Technical risk | Bias in AI | Racial bias | Science |
| 17 | Seyyed-Kalantari et al (2020) | Technical risk | Bias in AI | Racial bias | Proceedings of the Pacific Symposium. 2021 |
| 18 | Dusenberry (2018) | Technical risk | Bias in AI | Racial bias | The Health Gap, BBC News |
| 19 | Westergaard et al (2019) | Technical risk | Bias in AI | Racial bias | Nature communications |
| 20 | Rajkomar et al (2018) | Technical risk | Bias in AI | Racial bias | Annals of Internal Medicine |
| 21 | Kaushal et al (2020) | Technical risk | Bias in AI | Geographic bias | Jama |
| 22 | Wiggers (2020) | Socio-ethical risk | Lack of transparency | | VentureBeat |

| No | Authors | Risk Theme | Sub-Theme | Focus | Journal |
|---|---|---|---|---|---|
| 23 | Mora-Cantallops et al (2021) | Socio-ethical risk | Lack of transparency | Traceability | Big Data and Cognitive Computing |
| 24 | Selbst & Powles (2017) | Socio-ethical risk | Lack of transparency | Explainability | International Data Privacy Law |
| 25 | Yang et al (2021) | Socio-ethical risk | Lack of transparency | Explainability | Data Fusion: A Mini-Review, Two Showcases and Beyond |
| 26 | Lipton (2017) | Socio-ethical risk | Lack of transparency | Explainability | Independent article |
| 27 | Raji (2020) | Socio-ethical risk | Gaps in AI accountability | Legal gaps | Independent article |
| 28 | Smith (2020) | Socio-ethical risk | Gaps in AI accountability | Undefined Roles & responsibilities | AI & SOCIETY |
| 29 | Whitby (2015) | Socio-ethical risk | Gaps in AI accountability | Undefined Roles & responsibilities | Intelligent Systems, Control and Automation: Science and Engineering |
| 30 | Maliha et al (2021) | Socio-ethical risk | Gaps in AI accountability | Undefined Roles & responsibilities | The Milbank Quarterly |
| 31 | Harned et al (2019) | Socio-ethical risk | Gaps in AI accountability | Undefined Roles & responsibilities | Harvard Journal of Law & Technology |
| 32 | Pickering (2021) | Technical risk | Privacy and security issues | Data shared without consent | Future Internet |
| 33 | Ploug & Holm (2016) | Technical risk | Privacy and security issues | Data shared without consent | Bioethics |
| 34 | Vyas et al (2020) | Technical risk | Privacy and security issues | Data shared without consent | The New England Journal of Medicine |
| 35 | McKeown (2021) | Technical risk | Privacy and security issues | Data re-purposing | Science and Engineering Ethics |
| 36 | Koops (2021) | Technical risk | Privacy and security issues | Data re-purposing | Law, Innovation and Technology |
| 37 | Hocking et al (2019) | Technical risk | Privacy and security issues | Data re-purposing | RAND Corporation |
| 38 | Kiener (2020) | Technical risk | Privacy and security issues | Cyberattacks | The Conversation |
| 39 | Mulcahy (2021) | Technical risk | Privacy and security issues | Cyberattacks | U.S' WebMD |